# Stored Electromagnetic Field Energies in General Materials

Wen Geyi

Research Center of Applied Electromagnetics, Nanjing University of Information Science and Technology, Nanjing, Jiangsu, China 210044. Email:wgy@nuist.edu.cn.

The most general expressions of the stored energies for time-harmonic electromagnetic fields are derived from the time-domain Poynting theorem, and are valuable in characterizing the energy storage and transport properties of complex media. A new energy conservation law for the time-harmonic electromagnetic fields, which involves the derived general expressions of the stored energies, is introduced. In contrast to the well-established Poynting theorem for time-harmonic fields, the real part of the new energy conservation law gives an equation for the sum of stored electric and magnetic field energies; the imaginary part involves an equation related to the difference between the dissipated electric and magnetic field energies. In a lossless isotropic and homogeneous medium, the new energy conservation law has a clear physical implication: the stored electromagnetic field energy of a radiating system enclosed by a surface is equal to the total field energy inside the surface subtracted by the field energy flowing out of the surface.

## I. Introduction

In 1964[1], Ginzburg pointed out "Despite the fact that the problem of the conservation law and the expression for the energy density in electrodynamics is a fundamental one, there are certain aspects of it which have not yet been elucidated, in particular for the case of an absorbing dispersive medium." He also pointed out "When absorption is present, it is not in general possible to introduce phenomenologically the concept of the mean electromagnetic energy density." Ginzburg's notes revealed a longstanding problem in fundamental electrodynamics. Formulating the energy densities for time-harmonic fields in complex media has been an active research area for many years and tremendous efforts have been devoted to solving this problem [1]-[22]. A common understanding is that there are no general formulations for field energy densities that are valid for an arbitrary medium. Many expressions of energy densities for time-harmonic fields are derived from the following time-domain Poynting theorem [2]-[4]

$$-\mathbf{J}(\mathbf{r},t)\cdot\mathbf{E}(\mathbf{r},t) - \nabla\cdot\mathbf{S}(\mathbf{r},t) = \mathbf{E}(\mathbf{r},t)\cdot\frac{\partial\mathbf{D}(\mathbf{r},t)}{\partial t} + \mathbf{H}(\mathbf{r},t)\cdot\frac{\partial\mathbf{B}(\mathbf{r},t)}{\partial t}, \quad (1)$$

where $\mathbf{S}(\mathbf{r},t) = \mathbf{E}(\mathbf{r},t)\times\mathbf{H}(\mathbf{r},t)$ is the Poynting vector, while other notations in (1) have their conventional meanings. In this paper, we will use $f(\mathbf{r},t)$ to signify the field quantity in the time domain while its frequency-domain counterpart is denoted by $f(\mathbf{r},\omega)$. It is commonly believed that the right-hand side of (1) cannot be expressed as the time derivative of an instantaneous energy density for general time-dependent fields since most media are temporally dispersive [3][4]. It is also known that the right-hand side of (1) contains two distinct parts, the dissipation rate of the field energy and the rate of change of the stored field energy. To derive the energy densities for time-harmonic fields from (1) requires great care as different interpretations for the stored field energy and the dissipated energy may lead to conflicting results. As mentioned in [6], the expression for the energy density in a dispersive medium is not readily found in textbooks, and, as current literature indicates, there is no consensus on what is the correct expression. Another common understanding is that a detailed model at the microscopic level for the medium is needed when both dispersion and dissipation are present. This means that the energy density has to be formulated separately for every material [10]. There have been two different approaches in this regard. One is to combine the Poynting theorem (1) with the equations of motion for electric and magnetic polarizations, and the other is to use equivalent circuit models [6]-[19].

Understanding where and how EM energy is stored in a lossy material is interesting and is required in many applications related to material sciences (including the design of dielectrics for high density energy storage and nanomaterials for negative index and cloaking applications),

STORED ELECTROMAGNETIC FIELD ENERGIES IN GENERAL MATERIALS

which may range from direct current to optical frequencies[6][20][21]. In this paper, we limit our considerations to time-harmonic electromagnetic fields. We first demonstrate that the general expressions for the stored electric and magnetic field energies exist for an arbitrary medium and can be derived from the time-domain Poynting theorem. We then introduce a new energy conservation law for the time-harmonic EM fields, which contains the derived general expressions for the stored electric and magnetic field energies. The new energy conservation law gives two independent equations: one is for the sum of the stored electric and magnetic field energies and the other is the difference between the dissipated electric and magnetic field energies. The general field energy density expressions are applied to formulate the stored and dissipated energy densities in various complex media, which simplify the conventional approaches and help avoid erroneous and inaccurate results. We further show that, in a lossless isotropic and homogeneous medium, the new energy conservation law has a clear physical meaning: the stored electromagnetic field energy of a radiating system enclosed by a surface is equal to the total field energy inside the surface subtracted by the energy flowing out of the surface, thus yielding a natural definition for the stored energy around a radiating system.

## II. Stored and dissipated field energies in general materials

A longstanding problem in electromagnetism is finding the correct approach to partition the right-hand side of (1) into the rate of change of stored field energies and the dissipation rate of the field energy in a general medium. An attempt will be made in this section to solve this problem. In order to get some physical insights into the nature of the problem, let us first examine a typical RLC circuit shown in Figure 1. According to Kirchhoff's voltage law, we may write

$$v_{in} = iR + L\frac{di}{dt} + v_c.$$

where $v_c = \frac{1}{C}\int_{-\infty}^{t} idt$ denotes the voltage across the capacitor. The time-domain power balance relation for the RLC circuit can then be easily obtained as follows

$$v_{in}i = p_R + \frac{dw_L}{dt} + \frac{dw_C}{dt}, \qquad (2)$$

where $p_R = i^2 R$, $w_L = Li^2/2$, and $w_C = Cv_c^2/2$ are the dissipated power in the resistor $R$, the stored magnetic energy in the inductor $L$ and magnetic energy in the capacitor $C$ respectively. The left-hand side of (2) is the input power of the RLC circuit from the voltage source $v_{in}$, and the right-hand side of (2) denotes the rate of increase of energy in the RLC circuit, which can be decomposed into the sum of the rate of energy absorbed by the resistor, the rate of magnetic energy stored in the inductor and the rate of electric energy stored in the capacitor. In general, it is impossible to find an energy function $w_R$ such that the dissipated power can be written as the time derivative of $w_R$, i.e., $p_R = dw_R/dt$ since the integral $w_R = \int_{-\infty}^{t} i^2 R dt$ may not exist for an arbitrary time dependence. Equation (2) thus indicates that the input power of the RLC circuit is not always expressible as a complete differential of an energy function due to the heat loss.

Let us consider the time-domain Poynting theorem (1) in a source-free region ($\mathbf{J}(\mathbf{r},t) = 0$)

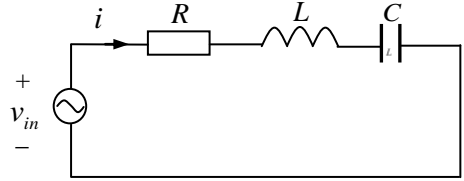

Figure 1 RLC circuit.

$$-\nabla \cdot \mathbf{S}(\mathbf{r},t) = \mathbf{E}(\mathbf{r},t) \cdot \frac{\partial \mathbf{D}(\mathbf{r},t)}{\partial t} + \mathbf{H}(\mathbf{r},t) \cdot \frac{\partial \mathbf{B}(\mathbf{r},t)}{\partial t}. \qquad (3)$$

Since the left-hand side of (3) represents the inflow power density from the external sources, the right-hand side of (3) must be interpreted as the sum of the dissipation rate of field energy density and the rate of increase of the total stored field energy density. Analogous to (2), the right-hand side of (3) may be decomposed into two parts. One part is converted into heat loss, and the other part is stored in the medium and is recoverable in the form of EM field energy, as described below

$$\mathbf{E}(\mathbf{r},t) \cdot \frac{\partial \mathbf{D}(\mathbf{r},t)}{\partial t} = p_{ed}(\mathbf{r},t) + \frac{dw_{es}(\mathbf{r},t)}{dt}, \qquad (4)$$

$$\mathbf{H}(\mathbf{r},t) \cdot \frac{\partial \mathbf{B}(\mathbf{r},t)}{\partial t} = p_{md}(\mathbf{r},t) + \frac{dw_{ms}(\mathbf{r},t)}{dt}, \qquad (5)$$

where $p_{ed}(\mathbf{r},t)$ and $p_{md}(\mathbf{r},t)$ denote the rate of dissipated electric and magnetic field energy density respectively, while $w_{es}(\mathbf{r},t)$ and $w_{ms}(\mathbf{r},t)$ stand for the stored electric and magnetic field energy density respectively. Accordingly (3) can be rewritten as

$$-\nabla \cdot \mathbf{S}(\mathbf{r},t) = p_{ed}(\mathbf{r},t) + p_{md}(\mathbf{r},t) + \frac{dw_{es}(\mathbf{r},t)}{dt} + \frac{dw_{ms}(\mathbf{r},t)}{dt}. \quad (6)$$

The above equation is similar to the power balance relation (2) for the RLC circuit. In order to validate (6), we only need to verify (4) and (5). For this purpose, we rewrite the left-hand sides of (4) and (5) as

$$\mathbf{E}(\mathbf{r},t) \cdot \frac{\partial \mathbf{D}(\mathbf{r},t)}{\partial t} = \frac{1}{2}\frac{d}{dt}\left[\mathbf{E}(\mathbf{r},t) \cdot \mathbf{D}(\mathbf{r},t)\right]$$
$$+ \frac{1}{2}\left[\mathbf{E}(\mathbf{r},t) \cdot \frac{\partial \mathbf{D}(\mathbf{r},t)}{\partial t} - \mathbf{D}(\mathbf{r},t) \cdot \frac{\partial \mathbf{E}(\mathbf{r},t)}{\partial t}\right], \quad (7)$$

$$\mathbf{H}(\mathbf{r},t) \cdot \frac{\partial \mathbf{B}(\mathbf{r},t)}{\partial t} = \frac{1}{2}\frac{d}{dt}\left[\mathbf{H}(\mathbf{r},t) \cdot \mathbf{B}(\mathbf{r},t)\right]$$
$$+ \frac{1}{2}\left[\mathbf{H}(\mathbf{r},t) \cdot \frac{\partial \mathbf{B}(\mathbf{r},t)}{\partial t} - \mathbf{B}(\mathbf{r},t) \cdot \frac{\partial \mathbf{H}(\mathbf{r},t)}{\partial t}\right]. \quad (8)$$

Since the stored field energies depend on the time evolution history of the system, we have to resort to the integration of (7) and (8) over the whole history of time evolution of the fields so that the stored energies can be identified. For this reason, we introduce two functions $u_e(\mathbf{r},t)$ and $u_m(\mathbf{r},t)$ respectively defined by

$$u_e(\mathbf{r},t) = \int_{-\infty}^{t} \mathbf{E}(\mathbf{r},t) \cdot \frac{\partial \mathbf{D}(\mathbf{r},t)}{\partial t} dt, \quad (9)$$

$$u_m(\mathbf{r},t) = \int_{-\infty}^{t} \mathbf{H}(\mathbf{r},t) \cdot \frac{\partial \mathbf{B}(\mathbf{r},t)}{\partial t} dt. \quad (10)$$

These integrals may not always exist for arbitrary time evolution. If the fields are assumed to be zero at $t = -\infty$, the general expressions for $u_e(\mathbf{r},t)$ and $u_m(\mathbf{r},t)$ can be easily found from (7) and (8) as follows

$$u_e(\mathbf{r},t) = \frac{1}{2}\mathbf{E}(\mathbf{r},t) \cdot \mathbf{D}(\mathbf{r},t)$$
$$+ \int_{-\infty}^{t} \frac{1}{2}\left[\mathbf{E}(\mathbf{r},t) \cdot \frac{\partial \mathbf{D}(\mathbf{r},t)}{\partial t} - \mathbf{D}(\mathbf{r},t) \cdot \frac{\partial \mathbf{E}(\mathbf{r},t)}{\partial t}\right] dt, \quad (11)$$

$$u_m(\mathbf{r},t) = \frac{1}{2}\mathbf{H}(\mathbf{r},t) \cdot \mathbf{B}(\mathbf{r},t)$$
$$+ \int_{-\infty}^{t} \frac{1}{2}\left[\mathbf{H}(\mathbf{r},t) \cdot \frac{\partial \mathbf{B}(\mathbf{r},t)}{\partial t} - \mathbf{B}(\mathbf{r},t) \cdot \frac{\partial \mathbf{H}(\mathbf{r},t)}{\partial t}\right] dt. \quad (12)$$

The above expressions, however, cannot directly be applied to time-harmonic fields since they are not zero at $t = -\infty$. To apply (11) and (12) to the time-harmonic fields, a damping factor (a small positive number) may first be introduced to the fields so that they die out at $t = -\infty$. The damping factor can then be sent to zero when the calculation is finished. This procedure seems reasonable as discussed in [22] since the time-harmonic fields in the strictest sense never occur in physical experiments. Let $\alpha$ denote the small damping factor. For the time-harmonic fields defined by

$$\mathbf{E}(\mathbf{r},t) = \mathrm{Re}\,\mathbf{E}(\mathbf{r},\omega)e^{j\omega t}, \text{ etc.} \quad (13)$$

where $\mathbf{E}(\mathbf{r},\omega)$ denotes the complex amplitude (phasor), we introduce the complex frequency $\tilde{\omega} = \omega - j\alpha$ to replace the real frequency $\omega$ in (13) to get

$$\tilde{\mathbf{E}}(\mathbf{r},t) = \mathrm{Re}\,\tilde{\mathbf{E}}(\mathbf{r},\tilde{\omega})e^{j\tilde{\omega}t}, \text{ etc.} \quad (14)$$

$\tilde{\mathbf{E}}(\mathbf{r},t)$ vanishes at $t = -\infty$ and reduces to $\mathbf{E}(\mathbf{r},t)$ as $\alpha$ approaches to zero. For sufficiently small $\alpha$ we can make the first-order approximations for the phasors

$$\tilde{\mathbf{E}}(\mathbf{r},\tilde{\omega}) \approx \mathbf{E}(\mathbf{r},\omega) - j\alpha \frac{\partial \mathbf{E}(\mathbf{r},\omega)}{\partial \omega}, \text{ etc.} \quad (15)$$

It follows from (14) and (15) that

$$\tilde{\mathbf{E}}(\mathbf{r},t) \cdot \tilde{\mathbf{D}}(\mathbf{r},t) \approx \frac{1}{2}\mathrm{Re}\left[\mathbf{E}(\mathbf{r},\omega) \cdot \mathbf{D}(\mathbf{r},\omega)e^{j2\omega t}\right]$$
$$+ \frac{1}{2}e^{2\alpha t}\mathrm{Re}\,\bar{\mathbf{E}}(\mathbf{r},\omega) \cdot \mathbf{D}(\mathbf{r},\omega) \quad (16)$$

$$\tilde{\mathbf{E}}(\mathbf{r},t) \cdot \frac{\partial \tilde{\mathbf{D}}(\mathbf{r},t)}{\partial t} - \tilde{\mathbf{D}}(\mathbf{r},t) \cdot \frac{\partial \tilde{\mathbf{E}}(\mathbf{r},t)}{\partial t}$$
$$\approx \omega e^{2\alpha t}\mathrm{Im}\,\mathbf{E}(\mathbf{r}) \cdot \bar{\mathbf{D}}(\mathbf{r}) \quad (17)$$
$$+ \alpha \omega e^{2\alpha t}\mathrm{Re}\left[\mathbf{E}(\mathbf{r},\omega) \cdot \frac{\partial \bar{\mathbf{D}}(\mathbf{r},\omega)}{\partial \omega} - \bar{\mathbf{D}}(\mathbf{r},\omega) \cdot \frac{\partial \mathbf{E}(\mathbf{r},\omega)}{\partial \omega}\right].$$

Inserting (16) and (17) into (11) gives

$$u_e(\mathbf{r},t) = \frac{e^{2\alpha t}}{4\alpha}\omega\,\mathrm{Im}\,\mathbf{E}(\mathbf{r}) \cdot \bar{\mathbf{D}}(\mathbf{r})$$
$$+ \frac{1}{4}\mathrm{Re}\left[\mathbf{E}(\mathbf{r},\omega) \cdot \mathbf{D}(\mathbf{r},\omega)e^{j2\omega t}\right]$$
$$+ \frac{1}{4}e^{2\alpha t}\mathrm{Re}\,\bar{\mathbf{E}}(\mathbf{r},\omega) \cdot \mathbf{D}(\mathbf{r},\omega) \quad (18)$$
$$+ \frac{e^{2\alpha t}}{4}\omega\,\mathrm{Re}\left[\mathbf{E}(\mathbf{r},\omega) \cdot \frac{\partial \bar{\mathbf{D}}(\mathbf{r},\omega)}{\partial \omega} - \bar{\mathbf{D}}(\mathbf{r},\omega) \cdot \frac{\partial \mathbf{E}(\mathbf{r},\omega)}{\partial \omega}\right].$$

After taking the derivative of $u_e(\mathbf{r},t)$ with respect to time and letting the damping constant $\alpha$ approach zero with the definition (9) taken into account, we immediately get

$$\mathbf{E}(\mathbf{r},t) \cdot \frac{\partial \mathbf{D}(\mathbf{r},t)}{\partial t} = \frac{\omega}{2}\mathrm{Im}\,\mathbf{E}(\mathbf{r},\omega) \cdot \bar{\mathbf{D}}(\mathbf{r},\omega)$$
$$+ \frac{d}{dt}\left\{\frac{1}{4}\mathrm{Re}\left[\mathbf{E}(\mathbf{r},\omega) \cdot \mathbf{D}(\mathbf{r},\omega)e^{j2\omega t}\right] + \frac{1}{4}\mathrm{Re}\,\bar{\mathbf{E}}(\mathbf{r},\omega) \cdot \mathbf{D}(\mathbf{r},\omega)\right.$$
$$\left. + \frac{1}{4}\omega\,\mathrm{Re}\left[\mathbf{E}(\mathbf{r},\omega) \cdot \frac{\partial \bar{\mathbf{D}}(\mathbf{r},\omega)}{\partial \omega} - \bar{\mathbf{D}}(\mathbf{r},\omega) \cdot \frac{\partial \mathbf{E}(\mathbf{r},\omega)}{\partial \omega}\right]\right\}.$$

The following identifications can be made if we compare the above equation with (4)

$$p_{ed}(\mathbf{r},t) = \frac{\omega}{2}\operatorname{Im}\mathbf{E}(\mathbf{r},\omega)\cdot\bar{\mathbf{D}}(\mathbf{r},\omega), \quad (19)$$

$$w_{es}(\mathbf{r},t) = \frac{1}{4}\operatorname{Re}\left[\mathbf{E}(\mathbf{r},\omega)\cdot\mathbf{D}(\mathbf{r},\omega)e^{j2\omega t}\right]$$
$$+ \frac{1}{4}\operatorname{Re}\bar{\mathbf{E}}(\mathbf{r},\omega)\cdot\mathbf{D}(\mathbf{r},\omega) \quad (20)$$
$$+ \frac{1}{4}\omega\operatorname{Re}\left[\mathbf{E}(\mathbf{r},\omega)\cdot\frac{\partial\bar{\mathbf{D}}(\mathbf{r},\omega)}{\partial\omega} - \bar{\mathbf{D}}(\mathbf{r},\omega)\cdot\frac{\partial\mathbf{E}(\mathbf{r},\omega)}{\partial\omega}\right].$$

Equation (5) can be obtained in a similar way. In this case, the magnetic field power loss density and stored magnetic field energy density can be identified as

$$p_{md}(\mathbf{r},t) = \frac{\omega}{2}\operatorname{Im}\left[\mathbf{H}(\mathbf{r},\omega)\cdot\bar{\mathbf{B}}(\mathbf{r},\omega)\right], \quad (21)$$

$$w_{ms}(\mathbf{r},t) = \frac{1}{4}\operatorname{Re}\left[\mathbf{H}(\mathbf{r},\omega)\cdot\mathbf{B}(\mathbf{r},\omega)e^{j2\omega t}\right]$$
$$+ \frac{1}{4}\operatorname{Re}\bar{\mathbf{H}}(\mathbf{r},\omega)\cdot\mathbf{B}(\mathbf{r},\omega) \quad (22)$$
$$+ \frac{1}{4}\omega\operatorname{Re}\left[\bar{\mathbf{H}}(\mathbf{r},\omega)\cdot\frac{\partial\mathbf{B}(\mathbf{r},\omega)}{\partial\omega} - \mathbf{B}(\mathbf{r},\omega)\cdot\frac{\partial\bar{\mathbf{H}}(\mathbf{r},\omega)}{\partial\omega}\right].$$

The (average) stored field energy densities and power loss density in the frequency domain can be found by taking the time averages of (19) to (22) over one period of the sinusoidal wave $e^{j\omega t}$ as follows

$$w_{es}(\mathbf{r},\omega) = \frac{1}{4}\operatorname{Re}\bar{\mathbf{E}}(\mathbf{r},\omega)\cdot\mathbf{D}(\mathbf{r},\omega)$$
$$+ \frac{1}{4}\omega\operatorname{Re}\left[\mathbf{E}(\mathbf{r},\omega)\cdot\frac{\partial\bar{\mathbf{D}}(\mathbf{r},\omega)}{\partial\omega} - \bar{\mathbf{D}}(\mathbf{r},\omega)\cdot\frac{\partial\mathbf{E}(\mathbf{r},\omega)}{\partial\omega}\right], \quad (23)$$

$$w_{ms}(\mathbf{r},\omega) = \frac{1}{4}\operatorname{Re}\bar{\mathbf{H}}(\mathbf{r},\omega)\cdot\mathbf{B}(\mathbf{r},\omega)$$
$$+ \frac{1}{4}\omega\operatorname{Re}\left[\bar{\mathbf{H}}(\mathbf{r},\omega)\cdot\frac{\partial\mathbf{B}(\mathbf{r},\omega)}{\partial\omega} - \mathbf{B}(\mathbf{r},\omega)\cdot\frac{\partial\bar{\mathbf{H}}(\mathbf{r},\omega)}{\partial\omega}\right], \quad (24)$$

$$p_{ed}(\mathbf{r},\omega) = \frac{\omega}{2}\operatorname{Im}\mathbf{E}(\mathbf{r},\omega)\cdot\bar{\mathbf{D}}(\mathbf{r},\omega), \quad (25)$$

$$p_{md}(\mathbf{r},\omega) = \frac{\omega}{2}\operatorname{Im}\left[\mathbf{H}(\mathbf{r},\omega)\cdot\bar{\mathbf{B}}(\mathbf{r},\omega)\right]. \quad (26)$$

The power loss densities (25) and (26) can also be derived from (1) using the method discussed in [3]. In fact, the time average of (3) over one period of the sinusoidal wave $e^{j\omega t}$ is easily shown to be

$$-\nabla\cdot\mathbf{S}(\mathbf{r},\omega) = \frac{\omega}{2}\operatorname{Im}\left[\mathbf{E}(\mathbf{r},\omega)\cdot\bar{\mathbf{D}}(\mathbf{r},\omega)\right]$$
$$+ \frac{\omega}{2}\operatorname{Im}\left[\mathbf{H}(\mathbf{r},\omega)\cdot\bar{\mathbf{B}}(\mathbf{r},\omega)\right]. \quad (27)$$

where $\mathbf{S}(\mathbf{r},\omega) = \frac{1}{2}\operatorname{Re}\left[\mathbf{E}(\mathbf{r},\omega)\times\bar{\mathbf{H}}(\mathbf{r},\omega)\right]$ is the time average of the time-domain Poynting vector $\mathbf{S}(\mathbf{r},t)$. The left-hand side of (27) denotes the inflow power density from external sources to maintain the field. Since the amplitude of the field is assumed to be constant in the steady state, all of the inflow power compensates the dissipation. For this reason, the right-hand side of (27) may be interpreted as the total dissipated power density of the medium. For a sinusoidal wave $e^{j\omega t}$, its period is given by $T = 2\pi/\omega$. The (average) dissipated electric and magnetic field energy densities can be defined as the (average) dissipated power multiplied by the period $T$

$$w_{ed}(\mathbf{r},\omega) = p_{ed}(\mathbf{r},\omega)T = \operatorname{Im}\pi\mathbf{E}(\mathbf{r},\omega)\cdot\bar{\mathbf{D}}(\mathbf{r},\omega), \quad (28)$$

$$w_{md}(\mathbf{r},\omega) = p_{md}(\mathbf{r},\omega)T = \operatorname{Im}\pi\mathbf{H}(\mathbf{r},\omega)\cdot\bar{\mathbf{B}}(\mathbf{r},\omega). \quad (29)$$

Note that the medium parameters do not explicitly appear in the energy density expressions (23) (24), (28) and (29). Therefore these expressions are applicable to any materials.

## III. New energy conservation law for time-harmonic fields

Once we have found the general expressions (23), (24), (28) and (29) for stored and dissipated field energies, an interesting question may be raised whether there exists an energy conservation law that involves these energy expressions. This question will be answered in this section by using a complex analysis. Applying the Laplace transform to the time-domain Maxwell equations

$$\nabla\times\mathbf{H}(\mathbf{r},t) = \mathbf{J}(\mathbf{r},t) + \frac{\partial\mathbf{D}(\mathbf{r},t)}{\partial t}, \nabla\times\mathbf{E}(\mathbf{r},t) = -\frac{\partial\mathbf{B}(\mathbf{r},t)}{\partial t},$$

we have

$$\begin{cases}\nabla\times\mathbf{H}(\mathbf{r},s) = \mathbf{J}(\mathbf{r},s) + s\mathbf{D}(\mathbf{r},s),\\ \nabla\times\mathbf{E}(\mathbf{r},s) = -s\mathbf{B}(\mathbf{r},s),\end{cases} \quad (30)$$

where $s = \alpha + j\omega$ denotes the complex frequency. Note that we use $f(\mathbf{r},s)$ to denote the field quantity in the complex frequency domain while the corresponding field quantity in the real frequency domain is denoted by $f(\mathbf{r},\omega)$. From (30), we obtain

$$\nabla\cdot\left[\mathbf{E}(\mathbf{r},s)\times\bar{\mathbf{H}}(\mathbf{r},s)\right] = -\mathbf{E}(\mathbf{r},s)\cdot\bar{\mathbf{J}}(\mathbf{r},s)$$
$$-\alpha\left[\bar{\mathbf{H}}(\mathbf{r},s)\cdot\mathbf{B}(\mathbf{r},s) + \mathbf{E}(\mathbf{r},s)\cdot\bar{\mathbf{D}}(\mathbf{r},s)\right] \quad (31)$$
$$-j\omega\left[\bar{\mathbf{H}}(\mathbf{r},s)\cdot\mathbf{B}(\mathbf{r},s) - \mathbf{E}(\mathbf{r},s)\cdot\bar{\mathbf{D}}(\mathbf{r},s)\right].$$

For an arbitrary analytic function $f(\mathbf{r},s)$, the Cauchy-Riemann conditions imply

$$\frac{\partial f(\mathbf{r},s)}{\partial \alpha} = -j\frac{\partial f(\mathbf{r},s)}{\partial \omega}. \tag{32}$$

If $\alpha$ is sufficiently small, we may have the following first order series expansion

$$f(\mathbf{r},s) \approx f(\mathbf{r},\omega) - j\alpha \frac{\partial f(\mathbf{r},\omega)}{\partial \omega}. \tag{33}$$

Using equations (32) and (33), equation (31) may be approximated by

$$\begin{aligned}
&\nabla \cdot \left[\mathbf{E}(\mathbf{r},\omega) \times \bar{\mathbf{H}}(\mathbf{r},\omega)\right] \\
&+ j\alpha \nabla \cdot \left[\mathbf{E}(\mathbf{r},\omega) \times \frac{\partial \bar{\mathbf{H}}(\mathbf{r},\omega)}{\partial \omega} - \frac{\partial \mathbf{E}(\mathbf{r},\omega)}{\partial \omega} \times \bar{\mathbf{H}}(\mathbf{r},\omega)\right] \\
&= -\mathbf{E}(\mathbf{r},\omega) \cdot \bar{\mathbf{J}}(\mathbf{r},\omega) \\
&- j\alpha \left[\mathbf{E}(\mathbf{r},\omega) \cdot \frac{\partial \bar{\mathbf{J}}(\mathbf{r},\omega)}{\partial \omega} - \frac{\partial \mathbf{E}(\mathbf{r},\omega)}{\partial \omega} \cdot \bar{\mathbf{J}}(\mathbf{r},\omega)\right] \\
&- j\omega \left[\mathbf{B}(\mathbf{r},\omega) \cdot \bar{\mathbf{H}}(\mathbf{r},\omega) - \mathbf{E}(\mathbf{r},\omega) \cdot \bar{\mathbf{D}}(\mathbf{r},\omega)\right] \\
&- \alpha \left[\mathbf{B}(\mathbf{r},\omega) \cdot \bar{\mathbf{H}}(\mathbf{r},\omega) + \mathbf{E}(\mathbf{r},\omega) \cdot \bar{\mathbf{D}}(\mathbf{r},\omega)\right] \\
&- \alpha\omega \left[\bar{\mathbf{H}}(\mathbf{r},\omega) \cdot \frac{\partial \mathbf{B}(\mathbf{r},\omega)}{\partial \omega} - \mathbf{B}(\mathbf{r},\omega) \cdot \frac{\partial \bar{\mathbf{H}}(\mathbf{r},\omega)}{\partial \omega}\right] \\
&- \alpha\omega \left[\mathbf{E}(\mathbf{r},\omega) \cdot \frac{\partial \bar{\mathbf{D}}(\mathbf{r},\omega)}{\partial \omega} - \bar{\mathbf{D}}(\mathbf{r},\omega) \cdot \frac{\partial \mathbf{E}(\mathbf{r},\omega)}{\partial \omega}\right].
\end{aligned} \tag{34}$$

Comparing the coefficients of similar terms, we obtain the well-known Poynting theorem for time-harmonic fields

$$\begin{aligned}
-\frac{1}{2}\mathbf{E}(\mathbf{r},\omega) \cdot \bar{\mathbf{J}}(\mathbf{r},\omega) &= \nabla \cdot \frac{1}{2}\left[\mathbf{E}(\mathbf{r},\omega) \times \bar{\mathbf{H}}(\mathbf{r},\omega)\right] \\
&+ j2\omega\left[\frac{1}{4}\mathbf{B}(\mathbf{r},\omega) \cdot \bar{\mathbf{H}}(\mathbf{r},\omega) - \frac{1}{4}\mathbf{E}(\mathbf{r},\omega) \cdot \bar{\mathbf{D}}(\mathbf{r},\omega)\right],
\end{aligned} \tag{35}$$

and the following new energy conservation law for an arbitrary medium

$$\begin{aligned}
&\frac{1}{4}\mathbf{E}(\mathbf{r},\omega) \cdot \bar{\mathbf{D}}(\mathbf{r},\omega) + \frac{1}{4}\mathbf{B}(\mathbf{r},\omega) \cdot \bar{\mathbf{H}}(\mathbf{r},\omega) \\
&+ \frac{1}{4}\omega\left[\mathbf{E}(\mathbf{r},\omega) \cdot \frac{\partial \bar{\mathbf{D}}(\mathbf{r},\omega)}{\partial \omega} - \bar{\mathbf{D}}(\mathbf{r},\omega) \cdot \frac{\partial \mathbf{E}(\mathbf{r},\omega)}{\partial \omega}\right] \\
&+ \frac{1}{4}\omega\left[\bar{\mathbf{H}}(\mathbf{r},\omega) \cdot \frac{\partial \mathbf{B}(\mathbf{r},\omega)}{\partial \omega} - \mathbf{B}(\mathbf{r},\omega) \cdot \frac{\partial \bar{\mathbf{H}}(\mathbf{r},\omega)}{\partial \omega}\right] \\
&= -j\frac{1}{4}\nabla \cdot \left[\mathbf{E}(\mathbf{r},\omega) \times \frac{\partial \bar{\mathbf{H}}(\mathbf{r},\omega)}{\partial \omega} - \frac{\partial \mathbf{E}(\mathbf{r},\omega)}{\partial \omega} \times \bar{\mathbf{H}}(\mathbf{r},\omega)\right] \\
&- j\frac{1}{4}\left[\mathbf{E}(\mathbf{r},\omega) \cdot \frac{\partial \bar{\mathbf{J}}(\mathbf{r},\omega)}{\partial \omega} - \frac{\partial \mathbf{E}(\mathbf{r},\omega)}{\partial \omega} \cdot \bar{\mathbf{J}}(\mathbf{r},\omega)\right].
\end{aligned} \tag{36}$$

Thus the complex analysis yields two energy conservation laws simultaneously. Equation (36) does not explicitly involve the medium parameters and can be verified by using the time-harmonic Maxwell equations (see Appendix). The physical significance of the new energy conservation law (36) becomes clear if it is decomposed into real and imaginary parts

$$\begin{aligned}
&w_{es}(\mathbf{r},\omega) + w_{ms}(\mathbf{r},\omega) = \\
&\nabla \cdot \mathrm{Im}\frac{1}{4}\left[\mathbf{E}(\mathbf{r},\omega) \times \frac{\partial \bar{\mathbf{H}}(\mathbf{r},\omega)}{\partial \omega} - \frac{\partial \mathbf{E}(\mathbf{r},\omega)}{\partial \omega} \times \bar{\mathbf{H}}(\mathbf{r},\omega)\right] \\
&+ \mathrm{Im}\frac{1}{4}\left[\mathbf{E}(\mathbf{r},\omega) \cdot \frac{\partial \bar{\mathbf{J}}(\mathbf{r},\omega)}{\partial \omega} - \frac{\partial \mathbf{E}(\mathbf{r},\omega)}{\partial \omega} \cdot \bar{\mathbf{J}}(\mathbf{r},\omega)\right],
\end{aligned} \tag{37}$$

$$\begin{aligned}
&w_{ed}(\mathbf{r},\omega) - w_{md}(\mathbf{r},\omega) = \\
&- \mathrm{Im}\,\pi\omega\left[\mathbf{E}(\mathbf{r},\omega) \cdot \frac{\partial \bar{\mathbf{D}}(\mathbf{r},\omega)}{\partial \omega} - \bar{\mathbf{D}}(\mathbf{r},\omega) \cdot \frac{\partial \mathbf{E}(\mathbf{r},\omega)}{\partial \omega}\right] \\
&- \mathrm{Im}\,\pi\omega\left[\bar{\mathbf{H}}(\mathbf{r},\omega) \cdot \frac{\partial \mathbf{B}(\mathbf{r},\omega)}{\partial \omega} - \mathbf{B}(\mathbf{r},\omega) \cdot \frac{\partial \bar{\mathbf{H}}(\mathbf{r},\omega)}{\partial \omega}\right] \\
&- \nabla \cdot \mathrm{Re}\,\pi\left[\mathbf{E}(\mathbf{r},\omega) \times \frac{\partial \bar{\mathbf{H}}(\mathbf{r},\omega)}{\partial \omega} - \frac{\partial \mathbf{E}(\mathbf{r},\omega)}{\partial \omega} \times \bar{\mathbf{H}}(\mathbf{r},\omega)\right] \\
&- \mathrm{Re}\,\pi\left[\mathbf{E}(\mathbf{r},\omega) \cdot \frac{\partial \bar{\mathbf{J}}(\mathbf{r},\omega)}{\partial \omega} - \frac{\partial \mathbf{E}(\mathbf{r},\omega)}{\partial \omega} \cdot \bar{\mathbf{J}}(\mathbf{r},\omega)\right],
\end{aligned} \tag{38}$$

where $w_{es}(\mathbf{r},\omega)$ and $w_{ms}(\mathbf{r},\omega)$ are defined by (23) and (24) while $w_{ed}(\mathbf{r},\omega)$ and $w_{md}(\mathbf{r},\omega)$ are defined by (28) and (29). As a result, the new energy conservation law gives two equations simultaneously, one is for the sum of stored electric and magnetic field energies and the other is for the difference of the dissipated electric and magnetic field energies, both being valid in an arbitrary medium. The stored field energy densities can be broken down into dominant parts

$$w_{es}^{dom}(\mathbf{r},\omega) = \mathrm{Re}\frac{1}{4}\mathbf{E}(\mathbf{r},\omega) \cdot \bar{\mathbf{D}}(\mathbf{r},\omega), \tag{39}$$

$$w_{ms}^{dom}(\mathbf{r},\omega) = \mathrm{Re}\frac{1}{4}\mathbf{B}(\mathbf{r},\omega) \cdot \bar{\mathbf{H}}(\mathbf{r},\omega), \tag{40}$$

and dispersive parts

$$\begin{aligned}
&w_{es}^{dis}(\mathbf{r},\omega) = \\
&\frac{1}{4}\mathrm{Re}\,\omega\left[\mathbf{E}(\mathbf{r},\omega) \cdot \frac{\partial \bar{\mathbf{D}}(\mathbf{r},\omega)}{\partial \omega} - \bar{\mathbf{D}}(\mathbf{r},\omega) \cdot \frac{\partial \mathbf{E}(\mathbf{r},\omega)}{\partial \omega}\right],
\end{aligned} \tag{41}$$

$$\begin{aligned}
&w_{ms}^{dis}(\mathbf{r},\omega) = \\
&\frac{1}{4}\mathrm{Re}\,\omega\left[\bar{\mathbf{H}}(\mathbf{r},\omega) \cdot \frac{\partial \mathbf{B}(\mathbf{r},\omega)}{\partial \omega} - \mathbf{B}(\mathbf{r},\omega) \cdot \frac{\partial \bar{\mathbf{H}}(\mathbf{r},\omega)}{\partial \omega}\right],
\end{aligned} \tag{42}$$

respectively. The dispersive parts of the stored field energies are caused by the dispersion of materials. We remark that some special forms of equation (36) already exist in previous studies for different purposes (e.g., [22]-[24]), but none of the authors realized their exact physical meaning as the general expressions for stored field energies were not available at the time.

STORED ELECTROMAGNETIC FIELD ENERGIES IN GENERAL MATERIALS

## IV. Stored and dissipated field energies in typical media

The new expressions (23), (24), (28) and (29) for stored and dissipated field energies provide straightforward formulations for the field energies in complex media, and will now be applied to typical media. It will be shown that the new stored energies in lossy materials contain additional terms that were missing in previous studies. These additional terms reflect the influences of the material losses on the stored energies.

### A. Isotropic media

Consider an isotropic medium defined by
$$\mathbf{D}(\mathbf{r},\omega) = \varepsilon(\mathbf{r},\omega)\mathbf{E}(\mathbf{r},\omega),\ \mathbf{B}(\mathbf{r},\omega) = \mu(\mathbf{r},\omega)\mathbf{H}(\mathbf{r},\omega),$$
with complex permittivity
$$\varepsilon(\mathbf{r},\omega) = \varepsilon'(\mathbf{r},\omega) - j\varepsilon''(\mathbf{r},\omega)$$
and complex permeability
$$\mu(\mathbf{r},\omega) = \mu'(\mathbf{r},\omega) - j\mu''(\mathbf{r},\omega)$$
where $\varepsilon'(\mathbf{r},\omega)$ and $\varepsilon''(\mathbf{r},\omega)$ are the capacity and dielectric loss factor; $\mu'(\mathbf{r},\omega)$ and $\mu''(\mathbf{r},\omega)$ are inductivity and magnetic loss factor. Note that losses due to conduction currents can be included in the complex permittivity. Substituting the medium parameters into (23), (24), (28) and (29), we have

$$w_{es}(\mathbf{r},\omega) = \frac{1}{4}\frac{\partial[\omega\varepsilon'(\mathbf{r},\omega)]}{\partial\omega}|\mathbf{E}(\mathbf{r},\omega)|^2 + \frac{\omega}{2}\varepsilon''(\mathbf{r},\omega)\,\text{Im}\left(\overline{\mathbf{E}}(\mathbf{r},\omega)\cdot\frac{\partial\mathbf{E}(\mathbf{r},\omega)}{\partial\omega}\right), \quad (43)$$

$$w_{ms}(\mathbf{r},\omega) = \frac{1}{4}\frac{\partial[\omega\mu'(\mathbf{r},\omega)]}{\partial\omega}|\mathbf{H}(\mathbf{r},\omega)|^2 + \frac{\omega}{2}\mu''(\mathbf{r},\omega)\,\text{Im}\,\overline{\mathbf{H}}(\mathbf{r},\omega)\cdot\frac{\partial\mathbf{H}(\mathbf{r},\omega)}{\partial\omega}, \quad (44)$$

$$w_{de}(\mathbf{r},\omega) = \pi\varepsilon''(\mathbf{r},\omega)|\mathbf{E}(\mathbf{r},\omega)|^2, \quad (45)$$

$$w_{dm}(\mathbf{r},\omega) = \pi\mu''(\mathbf{r},\omega)|\mathbf{H}(\mathbf{r},\omega)|^2. \quad (46)$$

It is interesting to note that both the stored electric and magnetic field energy densities in (43) and (44) contain frequency derivative terms, which are caused by the dielectric and magnetic losses and did not appear in previous reports (e.g., [3] and [4]). This is because stored field energy densities have usually been derived under an assumption that the fields are wave-packets (non-monochromatic) and their amplitudes vary very slowly with time and frequency as discussed in [3][4]. As a result, the frequency derivatives of the fields have been neglected. If losses can be neglected, (43) and (44) reduce to

$$w_{es}(\mathbf{r},\omega) = \frac{1}{4}\frac{\partial[\omega\varepsilon'(\mathbf{r},\omega)]}{\partial\omega}|\mathbf{E}(\mathbf{r},\omega)|^2, \quad (47)$$

$$w_{ms}(\mathbf{r},\omega) = \frac{1}{4}\frac{\partial[\omega\mu'(\mathbf{r},\omega)]}{\partial\omega}|\mathbf{H}(\mathbf{r},\omega)|^2. \quad (48)$$

These results are well established and their derivations are discussed in [3].

### B. Anisotropic media

For an anisotropic medium defined by
$$\mathbf{D}(\mathbf{r},\omega) = \ddot{\boldsymbol{\varepsilon}}(\mathbf{r},\omega)\cdot\mathbf{E}(\mathbf{r},\omega),\ \mathbf{B}(\mathbf{r},\omega) = \ddot{\boldsymbol{\mu}}(\mathbf{r},\omega)\cdot\mathbf{H}(\mathbf{r},\omega),$$
in which $\ddot{\boldsymbol{\varepsilon}}(\mathbf{r},\omega)$ and $\ddot{\boldsymbol{\mu}}(\mathbf{r},\omega)$ are dyads, the stored field energy densities can be written as

$$w_{es}(\mathbf{r},\omega) = \frac{1}{4}\text{Re}\,\mathbf{E}(\mathbf{r},\omega)\cdot\overline{\ddot{\boldsymbol{\varepsilon}}}(\mathbf{r},\omega)\cdot\overline{\mathbf{E}}(\mathbf{r},\omega)$$
$$+\frac{\omega}{4}\text{Re}\,\mathbf{E}(\mathbf{r},\omega)\cdot\frac{\partial\overline{\ddot{\boldsymbol{\varepsilon}}}(\mathbf{r},\omega)}{\partial\omega}\cdot\overline{\mathbf{E}}(\mathbf{r},\omega)$$
$$+\frac{\omega}{4}\text{Re}\,\overline{\mathbf{E}}(\mathbf{r},\omega)\cdot\ddot{\boldsymbol{\varepsilon}}(\mathbf{r},\omega)\cdot\frac{\partial\mathbf{E}(\mathbf{r},\omega)}{\partial\omega}$$
$$-\frac{\omega}{4}\text{Re}\,\overline{\mathbf{E}}(\mathbf{r},\omega)\cdot\ddot{\boldsymbol{\varepsilon}}^\dagger(\mathbf{r},\omega)\cdot\frac{\partial\mathbf{E}(\mathbf{r},\omega)}{\partial\omega},$$

$$w_{ms}(\mathbf{r},\omega) = \frac{1}{4}\text{Re}\,\overline{\mathbf{H}}(\mathbf{r},\omega)\cdot\ddot{\boldsymbol{\mu}}(\mathbf{r},\omega)\cdot\mathbf{H}(\mathbf{r},\omega)$$
$$+\frac{\omega}{4}\text{Re}\,\overline{\mathbf{H}}(\mathbf{r},\omega)\cdot\frac{\partial\ddot{\boldsymbol{\mu}}(\mathbf{r},\omega)}{\partial\omega}\cdot\mathbf{H}(\mathbf{r},\omega)$$
$$+\frac{\omega}{4}\text{Re}\,\overline{\mathbf{H}}(\mathbf{r},\omega)\cdot\ddot{\boldsymbol{\mu}}(\mathbf{r},\omega)\cdot\frac{\partial\mathbf{H}(\mathbf{r},\omega)}{\partial\omega}$$
$$-\frac{\omega}{4}\text{Re}\,\overline{\mathbf{H}}(\mathbf{r},\omega)\cdot\ddot{\boldsymbol{\mu}}^\dagger(\mathbf{r},\omega)\cdot\frac{\partial\mathbf{H}(\mathbf{r},\omega)}{\partial\omega},$$

where the superscript '$\dagger$' denotes the Hermitian transpose. For a lossless medium, we have $\ddot{\boldsymbol{\varepsilon}}(\mathbf{r},\omega) = \ddot{\boldsymbol{\varepsilon}}^\dagger(\mathbf{r},\omega)$ and $\ddot{\boldsymbol{\mu}}(\mathbf{r},\omega) = \ddot{\boldsymbol{\mu}}^\dagger(\mathbf{r},\omega)$ [4], and the stored energy densities reduce to

$$w_{es}(\mathbf{r},\omega) = \frac{1}{4}\overline{\mathbf{E}}(\mathbf{r},\omega)\cdot\frac{\partial[\omega\ddot{\boldsymbol{\varepsilon}}(\mathbf{r},\omega)]}{\partial\omega}\cdot\mathbf{E}(\mathbf{r},\omega), \quad (49)$$

$$w_{ms}(\mathbf{r},\omega) = \frac{1}{4}\overline{\mathbf{H}}(\mathbf{r},\omega)\cdot\frac{\partial[\omega\ddot{\boldsymbol{\mu}}(\mathbf{r},\omega)]}{\partial\omega}\cdot\mathbf{H}(\mathbf{r},\omega). \quad (50)$$

These expressions are well known and have been derived in [4] from (1) under the assumption that the fields are wave-packets and their amplitudes vary very slowly with time and frequency. It is worth mentioning that no additional assumption is needed in our derivation.

### C. Lorentz media

A Lorentz medium is characterized by the Lorentz model

$$\varepsilon(\omega) = \varepsilon_0 \left(1 + \frac{\omega_p^2}{\omega_0^2 - \omega^2 + j\gamma\omega}\right), \quad (51)$$

where $\omega_0$ is the resonant angular frequency of atoms (or structural elements) of material, $\omega_p$ is the characteristic frequency or plasma frequency, and $\gamma$ is the damping coefficient. Equation (51) is based on a simplified model of the structure of atoms, in which the nucleus is assumed to be much more massive than the electron. The electron is bound to the nucleus by a linear restoring force obeying Hooke's law and is obtained from the equation of motion of the electric polarization vector. By combining the Poynting theorem and the equation of motion of the electric polarization, it has been found that the stored electric field energy for the Lorentz mediun is of the form[12]

$$w_{es}(\mathbf{r},\omega) = \frac{1}{4}\varepsilon_0 \left[1 + \frac{\omega_p^2(\omega_0^2 + \omega^2)}{(\omega_0^2 - \omega^2)^2 + \gamma^2\omega^2}\right]|\mathbf{E}(\mathbf{r},\omega)|^2. \quad (52)$$

The above expression can also be obtained from the equivalent circuit model of a capacitor filled with the medium [13]. According to the new expression (43), the stored electric field energy density in a Lorentz medium is, however, given by

$$w_{es}(\mathbf{r},\omega) = \frac{1}{4}\varepsilon_0 \left[1 + \frac{\omega_p^2(\omega_0^2 + \omega^2)}{(\omega_0^2 - \omega^2)^2 + \gamma^2\omega^2}\right]|\mathbf{E}(\mathbf{r},\omega)|^2$$
$$- \frac{1}{2}\frac{\varepsilon_0 \omega^2 \omega_p^2 \gamma^2 (\omega^2 + \omega_0^2)}{\left[(\omega_0^2 - \omega^2)^2 + \gamma^2\omega^2\right]^2}|\mathbf{E}(\mathbf{r},\omega)|^2 \quad (53)$$
$$+ \frac{1}{2}\frac{\varepsilon_0 \omega^2 \omega_p^2 \gamma}{(\omega_0^2 - \omega^2)^2 + \gamma^2\omega^2} \text{Im}\left[\bar{\mathbf{E}}(\mathbf{r},\omega) \cdot \frac{\partial \mathbf{E}(\mathbf{r},\omega)}{\partial \omega}\right].$$

Notice that the last two terms on the right-hand side of (53) are missing in (52) although they are negligible for small $\gamma$.

Many artificial materials are characterized by the Lorentz model. Materials with negative permittivity can be realized by parallel thin metal wires [13]. If the wave vector is orthogonal to the wires while the electric field is parallel, the permittivity of the wire media and the corresponding stored field energy density can be respectively obtained from (51) and (53) by letting $\omega_0 = 0$.

### *D. General lossless media*

In a lossless medium, the process of building up an electromagnetic field is a reversible process. In this case, the right-hand side of (3) can be expressed as the time derivative of the total stored field energy density. A general lossless medium is defined by $\nabla \cdot \mathbf{S}(\mathbf{r},\omega) = 0$. It follows from time-harmonic Maxwell equations, we obtain

$$\nabla \cdot \mathbf{S}(\mathbf{r},\omega) = \nabla \cdot \frac{1}{2}\text{Re}\left[\mathbf{E}(\mathbf{r},\omega) \times \bar{\mathbf{H}}(\mathbf{r},\omega)\right]$$
$$= \frac{1}{2}\text{Re}\left\{j\omega\left[\mathbf{E}(\mathbf{r},\omega) \cdot \bar{\mathbf{D}}(\mathbf{r},\omega) - \bar{\mathbf{H}}(\mathbf{r},\omega) \cdot \mathbf{B}(\mathbf{r},\omega)\right]\right\}.$$

The lossless condition implies

$$\text{Im}\left[\mathbf{E}(\mathbf{r},\omega) \cdot \bar{\mathbf{D}}(\mathbf{r},\omega) - \bar{\mathbf{H}}(\mathbf{r},\omega) \cdot \mathbf{B}(\mathbf{r},\omega)\right]$$
$$= -\text{Im}\left[\bar{\mathbf{E}}(\mathbf{r},\omega) \cdot \mathbf{D}(\mathbf{r},\omega) + \bar{\mathbf{H}}(\mathbf{r},\omega) \cdot \mathbf{B}(\mathbf{r},\omega)\right] = 0. \quad (54)$$

in a source-free region. Taking the frequency derivative of (54) yields

$$\text{Im}\left[\bar{\mathbf{E}}(\mathbf{r},\omega) \cdot \frac{\partial \mathbf{D}(\mathbf{r},\omega)}{\partial \omega} + \mathbf{D}(\mathbf{r},\omega) \cdot \frac{\partial \bar{\mathbf{E}}(\mathbf{r},\omega)}{\partial \omega}\right]$$
$$+ \text{Im}\left[\bar{\mathbf{H}}(\mathbf{r},\omega) \cdot \frac{\partial \mathbf{B}(\mathbf{r},\omega)}{\partial \omega} + \mathbf{B}(\mathbf{r},\omega) \cdot \frac{\partial \bar{\mathbf{H}}(\mathbf{r},\omega)}{\partial \omega}\right]$$
$$= \text{Im}\left[\bar{\mathbf{E}}(\mathbf{r},\omega) \cdot \frac{\partial \mathbf{D}(\mathbf{r},\omega)}{\partial \omega} - \bar{\mathbf{D}}(\mathbf{r},\omega) \cdot \frac{\partial \mathbf{E}(\mathbf{r},\omega)}{\partial \omega}\right] \quad (55)$$
$$+ \text{Im}\left[\bar{\mathbf{H}}(\mathbf{r},\omega) \cdot \frac{\partial \mathbf{B}(\mathbf{r},\omega)}{\partial \omega} - \bar{\mathbf{B}}(\mathbf{r},\omega) \cdot \frac{\partial \mathbf{H}(\mathbf{r},\omega)}{\partial \omega}\right] = 0.$$

Using (54) and (55), we obtain

$$w_{es}(\mathbf{r},\omega) = \frac{1}{4}\bar{\mathbf{E}}(\mathbf{r},\omega) \cdot \mathbf{D}(\mathbf{r},\omega)$$
$$+ \frac{\omega}{4}\left[\bar{\mathbf{E}}(\mathbf{r},\omega) \cdot \frac{\partial \mathbf{D}(\mathbf{r},\omega)}{\partial \omega} - \bar{\mathbf{D}}(\mathbf{r},\omega) \cdot \frac{\partial \mathbf{E}(\mathbf{r},\omega)}{\partial \omega}\right], \quad (56)$$

$$w_{ms}(\mathbf{r},\omega) = \frac{1}{4}\mathbf{B}(\mathbf{r},\omega) \cdot \bar{\mathbf{H}}(\mathbf{r},\omega)$$
$$+ \frac{\omega}{4}\left[\bar{\mathbf{H}}(\mathbf{r},\omega) \cdot \frac{\partial \mathbf{B}(\mathbf{r},\omega)}{\partial \omega} - \bar{\mathbf{B}}(\mathbf{r},\omega) \cdot \frac{\partial \mathbf{H}(\mathbf{r},\omega)}{\partial \omega}\right]. \quad (57)$$

The stored field energy density expressions (56) and (57) agree with those obtained in [22], but have remained unnoticed for many years. They may be used to study the energy transport properties of a wavepacket and to prove that the energy velocity is always equal to the group velocity in a general lossless medium [25].

## V. Stored field energies around an electromagnetic radiator

The new energy conservation law (37) will now be applied to the study of the stored field energies for a radiating system. The study of the stored field energies of a radiating system is important and fundamental as they are closely related to the quality factor or the frequency bandwidth of the system. A radiating system is essentially equivalent to a RLC circuit, and the equivalent circuit parameters *R*, *L* and *C* are respectively determined by the radiated power, the stored magnetic field energy and the stored electric field energy. The stored field energy of a radiating system (not to be confused with the stored energy in materials discussed in Section II) is usually defined as the difference between the total EM field energy and the radiated energy, and has been investigated by many researchers (see [23]-[29] and references therein). Since both the total field energy and the radiated energy are infinite in an unbounded space for a time-harmonic field, their difference is expected to be a finite number. The above definition is actually based on an unproven hypothesis that the infinity in the total field energy is created by the energy flow associated with radiated power, which has raised discussions[27]. Indeed this definition has never been rigorously proven to be adequate; meanwhile, the new energy conservation law will be shown to be just right to fulfill the task. Let us consider an arbitrary radiator described by the source distribution $\mathbf{J}(\mathbf{r},\omega)$ confined in a finite region $V_0$ surrounded by a lossless isotropic homogeneous medium with permeability $\mu$ and permittivity $\varepsilon$, as shown in Figure 2. Taking the integration of (37) over the region $V$ bounded by a closed surface $S$ that contains $V_0$ and considering the fact that the dispersion energies (41) and (42) vanish in a lossless isotropic medium, we obtain

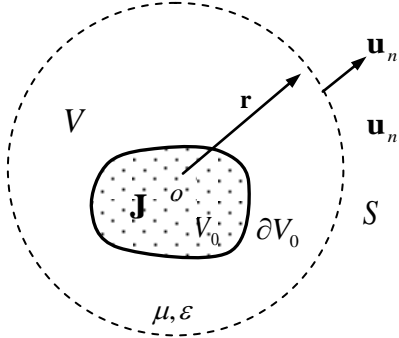

Figure 2 An arbitrary EM radiator.

$$\tilde{W}(\omega) = W_m^{dom}(\omega) + W_e^{dom}(\omega) - W_{rad}(\omega), \quad (58)$$

where $\tilde{W}(\omega)$ and $W_{rad}(\omega)$ are respectively defined by

$$\tilde{W}(\omega) = \frac{1}{4} \times$$
$$\mathrm{Im} \int_{V_0} \left[ \mathbf{E}(\mathbf{r},\omega) \cdot \frac{\partial \bar{\mathbf{J}}(\mathbf{r},\omega)}{\partial \omega} - \frac{\partial \mathbf{E}(\mathbf{r},\omega)}{\partial \omega} \cdot \bar{\mathbf{J}}(\mathbf{r},\omega) \right] dV(\mathbf{r}), \quad (59)$$

$$W_{rad}(\omega) = \frac{1}{4} \times$$
$$\mathrm{Im} \int_S \left[ \mathbf{E}(\mathbf{r},\omega) \times \frac{\partial \bar{\mathbf{H}}(\mathbf{r},\omega)}{\partial \omega} - \frac{\partial \mathbf{E}(\mathbf{r},\omega)}{\partial \omega} \times \bar{\mathbf{H}}(\mathbf{r},\omega) \right] \cdot \mathbf{u}_n(\mathbf{r}) dS(\mathbf{r}), \quad (60)$$

and

$$W_e^{dom}(\omega) = \frac{1}{4} \int_V \varepsilon |\mathbf{E}(\mathbf{r},\omega)|^2 dV(\mathbf{r}), \quad (61)$$

$$W_m^{dom}(\omega) = \frac{1}{4} \int_V \mu |\mathbf{H}(\mathbf{r},\omega)|^2 dV(\mathbf{r}) \quad (62)$$

respectively denote the total electric and magnetic field energy contained inside the region $V$. According to the energy balance relation (58), we may interpret $W_{rad}$ as the total EM field energy flowing out of the surface $S$ and interpret $\tilde{W}$ as the total stored EM energy of the radiator. Note that the total stored EM energy does not change as $S$ varies and is therefore a constant as long as the source region $V_0$ is included in $S$. By combining (58) and the imaginary part of the integral form of Poynting theorem (35) shown below

$$-\frac{1}{2} \mathrm{Im} \int_{V_0} \mathbf{E}(\mathbf{r},\omega) \cdot \bar{\mathbf{J}}(\mathbf{r},\omega) dV(\mathbf{r})$$
$$= \mathrm{Im} \int_S \frac{1}{2} \left[ \mathbf{E}(\mathbf{r},\omega) \times \bar{\mathbf{H}}(\mathbf{r},\omega) \right] \cdot \mathbf{u}_n(\mathbf{r}) dV(\mathbf{r}) \quad (63)$$
$$+ 2\omega \left[ W_m^{dom}(\omega) - W_e^{dom}(\omega) \right],$$

we may find

$$\tilde{W}_m(\omega) = W_m^{dom}(\omega) - \frac{1}{2} W_{rad}(\omega)$$
$$+ \frac{1}{8\omega} \mathrm{Im} \int_S \left[ \mathbf{E}(\mathbf{r},\omega) \times \bar{\mathbf{H}}(\mathbf{r},\omega) \right] \cdot \mathbf{u}_n(\mathbf{r}) dS(\mathbf{r}), \quad (64)$$

$$\tilde{W}_e(\omega) = W_e^{dom}(\omega) - \frac{1}{2} W_{rad}(\omega)$$
$$- \frac{1}{8\omega} \mathrm{Im} \int_S \left[ \mathbf{E}(\mathbf{r},\omega) \times \bar{\mathbf{H}}(\mathbf{r},\omega) \right] \cdot \mathbf{u}_n(\mathbf{r}) dS(\mathbf{r}). \quad (65)$$

Here $\tilde{W}_m(\omega)$ and $\tilde{W}_e(\omega)$ are defined by

$$\tilde{W}_m(\omega) = \frac{1}{8} \times$$
$$\mathrm{Im} \int_{V_0} \left[ \mathbf{E}(\mathbf{r},\omega) \cdot \frac{\partial \bar{\mathbf{J}}(\mathbf{r},\omega)}{\partial \omega} - \frac{\partial \mathbf{E}(\mathbf{r},\omega)}{\partial \omega} \cdot \bar{\mathbf{J}}(\mathbf{r},\omega) \right] dV(\mathbf{r}) \quad (66)$$
$$- \frac{1}{8\omega} \mathrm{Im} \int_{V_0} \mathbf{E}(\mathbf{r},\omega) \cdot \bar{\mathbf{J}}(\mathbf{r},\omega) dV(\mathbf{r}),$$

$$\tilde{W}_e(\omega) = \frac{1}{8} \times$$
$$\mathrm{Im} \int_{V_0} \left[ \mathbf{E}(\mathbf{r},\omega) \cdot \frac{\partial \bar{\mathbf{J}}(\mathbf{r},\omega)}{\partial \omega} - \frac{\partial \mathbf{E}(\mathbf{r},\omega)}{\partial \omega} \cdot \bar{\mathbf{J}}(\mathbf{r},\omega) \right] dV(\mathbf{r}) \quad (67)$$
$$+ \frac{1}{8\omega} \mathrm{Im} \int_{V_0} \mathbf{E}(\mathbf{r},\omega) \cdot \bar{\mathbf{J}}(\mathbf{r},\omega) dV(\mathbf{r}).$$

Note that the surface integral in (64) and (65) approaches zero as $S$ expands to infinity. From (64) and (65), we obtain

$$\begin{cases} \tilde{W}_m(\omega) + \tilde{W}_e(\omega) = \tilde{W}(\omega), \\ \tilde{W}_m(\omega) - \tilde{W}_e(\omega) = W_m^{dom}(\omega) - W_e^{dom}(\omega), \end{cases} \quad (68)$$

for infinitely large $S$. From the energy balance relation (58) and (68), we may interpret $\tilde{W}_m$ and $\tilde{W}_e$ as the stored magnetic field energy and stored electric field energy of the radiator. Note that both of them are independent of the choices of $S$ whenever the source is confined in $S$.

The integral (60) often occurs in the study of antenna input reactance [23][26][27], and can be simplified by using the following integral representations of EM fields[25]

$$\mathbf{E}(\mathbf{r},\omega) = -jk\eta \int_{V_0} G(R) \mathbf{J}(\mathbf{r}',\omega) dV(\mathbf{r}')$$
$$- \frac{\eta}{jk} \int_{V_0} \nabla' \cdot \mathbf{J}(\mathbf{r}',\omega) \nabla' G(R) dV(\mathbf{r}'), \quad (69)$$

$$\mathbf{H}(\mathbf{r},\omega) = \int_{V_0} \mathbf{J}(\mathbf{r}',\omega) \times \nabla' G(R) dV(\mathbf{r}'), \quad (70)$$

where $G(R) = e^{-jkR}/4\pi R$ is the fundamental solution of Helmholtz equation with $R = |\mathbf{r} - \mathbf{r}'|$; $\eta = \sqrt{\mu/\varepsilon}$ is the wave impedance; and $k = \omega\sqrt{\mu\varepsilon}$ is the wave number. Assume that the surface $S$ is a sufficiently large sphere of radius $r = |\mathbf{r}|$. As $r$ approaches to infinity, we have $\mathbf{u}_n(\mathbf{r}) \approx \mathbf{u}_r \equiv \mathbf{r}/r$ and the far-field relation

$$\mathbf{u}_r \times \mathbf{E}(\mathbf{r},\omega) \approx \eta \mathbf{H}(\mathbf{r},\omega)$$

on the spherical surface $S$. Taking the frequency derivatives of (69) and (70) first and then using the far-field approximations, we may find that

$$\mathbf{u}_r \times \frac{\partial \mathbf{E}(\mathbf{r},\omega)}{\partial \omega} \approx \eta \frac{\partial \mathbf{H}(\mathbf{r},\omega)}{\partial \omega}. \quad (71)$$

As a result, we have

$$\frac{1}{4}\left( \mathbf{E}(\mathbf{r},\omega) \times \frac{\partial \bar{\mathbf{H}}(\mathbf{r},\omega)}{\partial \omega} - \frac{\partial \mathbf{E}(\mathbf{r},\omega)}{\partial \omega} \times \bar{\mathbf{H}}(\mathbf{r},\omega) \right) \cdot \mathbf{u}_r \quad (72)$$
$$\approx -\frac{\eta}{2} \frac{\partial \mathbf{H}(\mathbf{r},\omega)}{\partial \omega} \cdot \bar{\mathbf{H}}(\mathbf{r},\omega).$$

Traditionally the right-hand side of (72) is further simplified by introducing the far-field approximation for the magnetic field $\mathbf{H}(\mathbf{r},\omega) = \mathbf{H}_\infty(\mathbf{u}_r,\omega)e^{-jkr}/r$ so that (60) can be expressed as

$$W_{rad}(\omega) = \frac{r}{c} P_{rad}(\omega)$$
$$- \frac{\eta}{2} \int_S \mathrm{Im}\left[ \bar{\mathbf{H}}_\infty(\mathbf{u}_r,\omega) \cdot \frac{\partial \mathbf{H}_\infty(\mathbf{u}_r,\omega)}{\partial \omega} \right] d\Omega(\mathbf{u}_r), \quad (73)$$

where $d\Omega(\mathbf{u}_r) = \sin\theta d\theta d\varphi$ is the differential solid angle; $P_{rad}(\omega) = \frac{1}{2} \int_S \eta |\mathbf{H}(\mathbf{r},\omega)|^2 dS(\mathbf{r})$ is the total radiated power and $\mathbf{H}_\infty(\mathbf{u}_r,\omega)$ is the magnetic far-field pattern. The physical meaning of the second term on the right-hand side of (73) has never been identified and has bewildered researchers for many years. It is this term that has caused a coordinate-dependent term in the expressions for the stored energies [23], which is unacceptable physically. In the following, we will demonstrate that this term is related to the frequency derivative of the source and is therefore negligible for small antenna. Equation (73) has been obtained from (72) by directly applying the frequency derivative to the far-field expression. A more accurate computation for the frequency derivative term on the right-hand side of (72) will be introduced below. Instead of applying the frequency derivative directly to the far-field expression for the magnetic field, we first take the frequency derivative of the exact magnetic field expression and then make a far-field approximation to the frequency derivative of the magnetic field. In the far-field region, the magnetic field and its frequency derivative are given by [25]

$$\mathbf{H}(\mathbf{r},\omega) \approx \frac{jk}{4\pi} \frac{e^{-jkr}}{r} \int_{V_0} [\mathbf{J}(\mathbf{r}',\omega) \times \mathbf{u}_r] e^{jk\mathbf{u}_r \cdot \mathbf{r}'} dV(\mathbf{r}'), \quad (74)$$

$$\frac{\partial \mathbf{H}(\mathbf{r},\omega)}{\partial \omega} \approx -j\frac{r}{c} \mathbf{H}(\mathbf{r},\omega)$$
$$+ \frac{jk}{4\pi} \frac{e^{-jkr}}{r} \int_{V_0} \left[ \frac{\partial \mathbf{J}(\mathbf{r}',\omega)}{\partial \omega} \times \mathbf{u}_r \right] e^{jk\mathbf{u}_r \cdot \mathbf{r}'} dV(\mathbf{r}'). \quad (75)$$

Notice that the frequency derivative (75) is obtained from the far-field approximation of the following exact frequency derivative for the magnetic field

$$\frac{\partial \mathbf{H}(\mathbf{r},\omega)}{\partial \omega} = \frac{1}{4\pi} \int_{V_0} \frac{\partial \mathbf{J}(\mathbf{r}',\omega)}{\partial \omega} \times \nabla' \frac{e^{-jkR}}{R} dV(\mathbf{r}')$$
$$- \frac{j}{4\pi c} \int_{V_0} \mathbf{J}(\mathbf{r}',\omega) \times \nabla' e^{-jkR} dV(\mathbf{r}'),$$

and is more accurate than that obtained from the frequency derivative of the far-field approximation (74). Thus

$$\frac{\partial \mathbf{H}(\mathbf{r},\omega)}{\partial \omega} \cdot \bar{\mathbf{H}}(\mathbf{r},\omega) = -j\frac{r}{c} |\mathbf{H}(\mathbf{r},\omega)|^2 + \frac{k^2}{16\pi^2 r^2} \times$$
$$\int_{V_0}\int_{V_0} \left[ \bar{\mathbf{J}}(\mathbf{r}',\omega) \cdot (\ddot{\mathbf{I}} - \mathbf{u}_r \mathbf{u}_r) \cdot \frac{\partial \mathbf{J}(\mathbf{r}'',\omega)}{\partial \omega} \right] e^{j k \mathbf{u}_r \cdot \mathbf{R}} dV(\mathbf{r}') dV(\mathbf{r}''),$$
(76)

with $\mathbf{R} = \mathbf{r}'' - \mathbf{r}'$, and (72) can be rewritten as

$$\frac{1}{4}\int_S \left[ \mathbf{E}(\mathbf{r},\omega) \times \frac{\partial \bar{\mathbf{H}}(\mathbf{r},\omega)}{\partial \omega} - \frac{\partial \mathbf{E}(\mathbf{r},\omega)}{\partial \omega} \times \bar{\mathbf{H}}(\mathbf{r},\omega) \right] \cdot$$
$$\mathbf{u}_n(\mathbf{r}) dS(\mathbf{r}) = j\frac{r}{c}\int_S \frac{\eta}{2}|\mathbf{H}(\mathbf{r},\omega)|^2 dS \quad (77)$$
$$- \frac{k^2\eta}{8\pi} \int_{V_0}\int_{V_0} \left[ \bar{\mathbf{J}}(\mathbf{r}',\omega) \cdot \ddot{\mathbf{U}} \cdot \frac{\partial \mathbf{J}(\mathbf{r}'',\omega)}{\partial \omega} \right] dV(\mathbf{r}') dV(\mathbf{r}''),$$

where $\ddot{\mathbf{U}}$ is a dyad defined by

$$\ddot{\mathbf{U}} = \frac{1}{4\pi} \int_S \frac{1}{r^2} (\ddot{\mathbf{I}} - \mathbf{u}_r \mathbf{u}_r) e^{j k \mathbf{u}_r \cdot \mathbf{R}} dS(\mathbf{r}), \quad (78)$$

with $\ddot{\mathbf{I}}$ being the unit dyad. The above integral can be carried out by the Funk-Hecke formula in the theory of spherical harmonics [30] (see also [28])

$$\ddot{\mathbf{U}} = \left[ j_0(kR) - \frac{j_1(kR)}{kR} \right] \ddot{\mathbf{I}} + j_2(kR) \mathbf{u}_R \mathbf{u}_R, \quad (79)$$

where $R = |\mathbf{r}' - \mathbf{r}''|$, $j_n$ is the $n$th-order spherical Bessel function of the first kind [31]. Substituting (77) into (60), we obtain

$$W_{rad}(\omega) = \frac{r}{c} P_{rad}(\omega) - W_{fd}(\omega), \quad (80)$$

where $W_{fd}(\omega)$ stands for the energy term caused by the frequency derivative of the source

$$W_{fd}(\omega) =$$
$$\frac{k^2\eta}{8\pi} \mathrm{Im} \int_{V_0}\int_{V_0} \left[ \bar{\mathbf{J}}(\mathbf{r}',\omega) \cdot \ddot{\mathbf{U}} \cdot \frac{\partial \mathbf{J}(\mathbf{r}'',\omega)}{\partial \omega} \right] dV(\mathbf{r}') dV(\mathbf{r}''). \quad (81)$$

Substituting (79) into (81) and simplifying the resultant yield

$$W_{fd}(\omega) = -\frac{k^2 c^2 \eta}{8\pi} \times$$
$$\int_{V_0}\int_{V_0} \mathrm{Im}\left[ \bar{\rho}(\mathbf{r}',\omega) \frac{\partial \rho(\mathbf{r}'',\omega)}{\partial \omega} \right] j_0(kR) dV(\mathbf{r}') dV(\mathbf{r}'') \quad (82)$$
$$+ \frac{k^2 \eta}{8\pi} \int_{V_0}\int_{V_0} \mathrm{Im}\left[ \bar{\mathbf{J}}(\mathbf{r}',\omega) \cdot \frac{\partial \mathbf{J}(\mathbf{r}'',\omega)}{\partial \omega} \right] j_0(kR) dV(\mathbf{r}') dV(\mathbf{r}''),$$

where $\rho(\mathbf{r},\omega) = j\nabla \cdot \mathbf{J}(\mathbf{r},\omega)/\omega$ is the charge density.

Comparing (73) with (80), we may find that the energy term $W_{fd}$ is approximately equal to the second term on the right-hand side of (73), i.e.,

$$W_{fd}(\omega) \approx \frac{\eta}{2} \int_S \mathrm{Im}\left[ \bar{\mathbf{H}}_\infty(\mathbf{u}_r,\omega) \cdot \frac{\partial \mathbf{H}_\infty(\mathbf{u}_r,\omega)}{\partial \omega} \right] d\Omega(\mathbf{u}_r). \quad (83)$$

If the source distribution is a slowly varying function of frequency (e.g. for a small antenna), the energy term $W_{fd}(\omega)$ can be neglected according to (81) or (82). As a result, the right-hand side of (83) is also negligible in this case.

Since the energy term $W_{fd}(\omega)$ does not depend on the coordinate system, the coordinate-dependent term arising from the far-field pattern term on the right-hand side of (73) is actually caused by the inaccurate approximation that the frequency derivative is directly applied to the far-field pattern. Substituting (80) into (58) yields

$$\tilde{W}(\omega) = W_m^{dom}(\omega) + W_e^{dom}(\omega) - \frac{r}{c} P_{rad}(\omega) + W_{fd}(\omega). \quad (84)$$

Since $W_{fd}(\omega)$ is negligible for small antenna, $\tilde{W}(\omega)$ reduces to the old definition of the stored energy for small antenna, denoted $\tilde{W}_{old}(\omega)$, defined by [26]

$$\tilde{W}_{old}(\omega) = W_m^{dom}(\omega) + W_e^{dom}(\omega) - \frac{r}{c} P_{rad}(\omega)$$
$$= \tilde{W}(\omega) - W_{fd}(\omega), \quad (85)$$

which is also independent of the coordinate system since both $\tilde{W}(\omega)$ and $W_{fd}(\omega)$ are independent of coordinate system. It should be mentioned that (58) has also been derived in [24], where the frequency derivative $\partial \bar{\mathbf{J}}(\mathbf{r},\omega)/\partial \omega$ has been set to zero by the author without going one step further to investigate the physical meaning of the equation.

The new definitions (59), (66) and (67) for stored field energies can be expressed in terms of the source distributions only. Using the integral representation for the electric field (69) and the Green's identity $\nabla \cdot (\phi \mathbf{a}) = \mathbf{a} \cdot \nabla \phi + \phi \nabla \cdot \mathbf{a}$, we may find

$$\mathrm{Im} \int_{V_0} \mathbf{E}(\mathbf{r},\omega) \cdot \frac{\partial \overline{\mathbf{J}}(\mathbf{r},\omega)}{\partial \omega} dV(\mathbf{r}) =$$

$$\frac{\eta c}{4\pi} \int_{V_0}\int_{V_0} \overline{\rho}(\mathbf{r},\omega)\rho(\mathbf{r}',\omega) \frac{\cos kR}{R} dV(\mathbf{r})dV(\mathbf{r}')$$

$$+\omega\eta c\, \mathrm{Im} \int_{V_0}\int_{V_0} j\frac{\partial \overline{\rho}(\mathbf{r},\omega)}{\partial \omega}\rho(\mathbf{r}',\omega)G(R)dV(\mathbf{r})dV(\mathbf{r}')$$

$$-\frac{\omega\eta}{c}\,\mathrm{Im}\int_{V_0}\int_{V_0} jG(R)\mathbf{J}(\mathbf{r}',\omega)\cdot\frac{\partial \overline{\mathbf{J}}(\mathbf{r},\omega)}{\partial \omega}dV(\mathbf{r})dV(\mathbf{r}'),$$
(86)

$$\mathrm{Im}\int_V \overline{\mathbf{J}}(\mathbf{r},\omega)\cdot\frac{\partial \mathbf{E}(\mathbf{r},\omega)}{\partial \omega}dV(\mathbf{r}) =$$

$$\omega\eta c\,\mathrm{Im}\int_V\int_V j\overline{\rho}(\mathbf{r},\omega)\frac{\partial \rho(\mathbf{r}',\omega)}{\partial \omega}G(R)dV(\mathbf{r})dV(\mathbf{r}')$$

$$+\omega\eta\,\mathrm{Im}\int_V\int_V \overline{\rho}(\mathbf{r},\omega)\rho(\mathbf{r}',\omega)RG(R)dV(\mathbf{r})dV(\mathbf{r}')$$

$$-\frac{\eta}{c}\,\mathrm{Im}\int_V\int_V j\overline{\mathbf{J}}(\mathbf{r},\omega)\cdot\mathbf{J}(\mathbf{r}',\omega)G(R)dV(\mathbf{r})dV(\mathbf{r}')$$

$$-\frac{\omega\eta}{c}\,\mathrm{Im}\int_V\int_V j\overline{\mathbf{J}}(\mathbf{r},\omega)\cdot\frac{\partial \mathbf{J}(\mathbf{r}',\omega)}{\partial \omega}G(R)dV(\mathbf{r})dV(\mathbf{r}')$$

$$-\frac{\omega\eta}{c^2}\,\mathrm{Im}\int_V\int_V \overline{\mathbf{J}}(\mathbf{r},\omega)\cdot\mathbf{J}(\mathbf{r}',\omega)RG(R)dV(\mathbf{r})dV(\mathbf{r}').$$
(87)

Substituting (86) and (87) into (66) and (67), and omitting the tedious derivation process, the stored field energies $\tilde{W}_e(\omega)$ and $\tilde{W}_m(\omega)$ are found to be

$$\tilde{W}_e(\omega) = \frac{\eta c}{16\pi} \int_{V_0}\int_{V_0} \overline{\rho}(\mathbf{r},\omega)\rho(\mathbf{r}',\omega)\frac{\cos kR}{R}dV(\mathbf{r})dV(\mathbf{r}')$$

$$+\frac{\omega\eta}{32\pi}\int_{V_0}\int_{V_0}\left[\overline{\rho}(\mathbf{r},\omega)\rho(\mathbf{r}') - \frac{1}{c^2}\overline{\mathbf{J}}(\mathbf{r},\omega)\cdot\mathbf{J}(\mathbf{r}',\omega)\right]\times$$

$$\sin kR\, dV(\mathbf{r})dV(\mathbf{r}')$$

$$-\frac{\omega\eta c}{16\pi}\int_{V_0}\int_{V_0}\mathrm{Im}\left[\overline{\rho}(\mathbf{r},\omega)\frac{\partial \rho(\mathbf{r}',\omega)}{\partial \omega}\right]\frac{\sin kR}{R}dV(\mathbf{r})dV(\mathbf{r}')$$

$$+\frac{\omega\eta c}{16\pi}\int_{V_0}\int_{V_0}\frac{1}{c^2}\mathrm{Im}\left[\overline{\mathbf{J}}(\mathbf{r},\omega)\cdot\frac{\partial \mathbf{J}(\mathbf{r}',\omega)}{\partial \omega}\right]\frac{\sin kR}{R}dV(\mathbf{r})dV(\mathbf{r}'),$$
(88)

$$\tilde{W}_m(\omega) = \frac{\eta c}{16\pi}\int_{V_0}\int_{V_0}\frac{1}{v^2}\overline{\mathbf{J}}(\mathbf{r},\omega)\cdot\mathbf{J}(\mathbf{r}',\omega)\frac{\cos kR}{R}dV(\mathbf{r})dV(\mathbf{r}')$$

$$+\frac{\omega\eta}{32\pi}\int_{V_0}\int_{V_0}\left[\overline{\rho}(\mathbf{r},\omega)\rho(\mathbf{r}',\omega) - \frac{1}{v^2}\overline{\mathbf{J}}(\mathbf{r},\omega)\cdot\mathbf{J}(\mathbf{r}',\omega)\right]\times$$

$$\sin kR\, dV(\mathbf{r})dV(\mathbf{r}')$$

$$-\frac{\omega\eta c}{16\pi}\int_{V_0}\int_{V_0}\mathrm{Im}\left[\overline{\rho}(\mathbf{r},\omega)\frac{\partial \rho(\mathbf{r}',\omega)}{\partial \omega}\right]\frac{\sin kR}{R}dV(\mathbf{r})dV(\mathbf{r}')$$

$$+\frac{\omega\eta v}{16\pi}\int_{V_0}\int_{V_0}\frac{1}{c^2}\mathrm{Im}\left[\overline{\mathbf{J}}(\mathbf{r},\omega)\cdot\frac{\partial \mathbf{J}(\mathbf{r}',\omega)}{\partial \omega}\right]\frac{\sin kR}{R}dV(\mathbf{r})dV(\mathbf{r}').$$
(89)

The expressions (88) and (89) agree with those obtained in [29], where a different approach has been adopted.

## VI. Conclusion

In this paper, three different but related problems have been investigated. Firstly, the longstanding problem of evaluating the stored electric and magnetic field energies in general media has been solved, and the general expressions for the stored energies have been obtained and validated by a number of applications. These general expressions for the stored energies facilitate the formulation of the stored energy densities in various complex media and help avoid the conflicting or erroneous results due to the difficulties in identifying the stored field energy and the dissipated energy in conventional approaches. Secondly, a new energy conservation law for time-harmonic fields in an arbitrary medium has been derived, which involves the general stored energy expressions found in this paper. Thirdly, the new energy conservation law has been shown to provide a natural definition for the stored energies around a radiating system. We believe that the results presented here are of fundamental importance and will have a positive impact on the study of energy storage and transport properties in various new materials for electrical and optical engineering. In this respect, much remains to be explored.

## Appendix: An alternative derivation of new energy conservation law

An alternative derivation for the new energy conservation law (36) will be presented below, which is also helpful for pedagogical purposes. We first take the complex conjugate of the first equation of (30) in the real frequency domain (i.e., $\alpha = 0$), and then take the frequency derivative of both the resultant equation and the second equation of (30), and the results are

$$\begin{cases} \nabla \times \dfrac{\partial \bar{\mathbf{H}}(\mathbf{r},\omega)}{\partial \omega} = \dfrac{\partial \bar{\mathbf{J}}(\mathbf{r},\omega)}{\partial \omega} - j\bar{\mathbf{D}}(\mathbf{r},\omega) - j\omega \dfrac{\partial \bar{\mathbf{D}}(\mathbf{r},\omega)}{\partial \omega}, \\ \nabla \times \dfrac{\partial \mathbf{E}(\mathbf{r},\omega)}{\partial \omega} = -j\mathbf{B}(\mathbf{r},\omega) - j\omega \dfrac{\partial \mathbf{B}(\mathbf{r},\omega)}{\partial \omega}. \end{cases}$$

Multiplying the first equation by $\mathbf{E}$ and the second by $\bar{\mathbf{H}}$ and adding the resultant equations yield

$$\mathbf{E}(\mathbf{r},\omega) \cdot \nabla \times \dfrac{\partial \bar{\mathbf{H}}(\mathbf{r},\omega)}{\partial \omega} + \bar{\mathbf{H}}(\mathbf{r},\omega) \cdot \nabla \times \dfrac{\partial \mathbf{E}(\mathbf{r},\omega)}{\partial \omega}$$
$$= \mathbf{E}(\mathbf{r},\omega) \cdot \dfrac{\partial \bar{\mathbf{J}}(\mathbf{r},\omega)}{\partial \omega} - j\mathbf{E}(\mathbf{r},\omega) \cdot \bar{\mathbf{D}}(\mathbf{r},\omega) - j\bar{\mathbf{H}}(\mathbf{r},\omega) \cdot \mathbf{B}(\mathbf{r},\omega)$$
$$- j\omega \mathbf{E}(\mathbf{r},\omega) \cdot \dfrac{\partial \bar{\mathbf{D}}(\mathbf{r},\omega)}{\partial \omega} - j\omega \bar{\mathbf{H}}(\mathbf{r},\omega) \cdot \dfrac{\partial \mathbf{B}(\mathbf{r},\omega)}{\partial \omega}.$$

This can be rearranged as

$$j\mathbf{E}(\mathbf{r},\omega) \cdot \bar{\mathbf{D}}(\mathbf{r},\omega) + j\bar{\mathbf{H}}(\mathbf{r},\omega) \cdot \mathbf{B}(\mathbf{r},\omega) = \mathbf{E}(\mathbf{r},\omega) \cdot \dfrac{\partial \bar{\mathbf{J}}(\mathbf{r},\omega)}{\partial \omega}$$
$$-\mathbf{E}(\mathbf{r},\omega) \cdot \nabla \times \dfrac{\partial \bar{\mathbf{H}}(\mathbf{r},\omega)}{\partial \omega} - \bar{\mathbf{H}}(\mathbf{r},\omega) \cdot \nabla \times \dfrac{\partial \mathbf{E}(\mathbf{r},\omega)}{\partial \omega}$$
$$-j\omega \left[ \mathbf{E}(\mathbf{r},\omega) \cdot \dfrac{\partial \bar{\mathbf{D}}(\mathbf{r},\omega)}{\partial \omega} - \dfrac{\partial \mathbf{E}(\mathbf{r},\omega)}{\partial \omega} \cdot \bar{\mathbf{D}}(\mathbf{r},\omega) \right]$$
$$-j\omega \left[ \bar{\mathbf{H}}(\mathbf{r},\omega) \cdot \dfrac{\partial \mathbf{B}(\mathbf{r},\omega)}{\partial \omega} - \dfrac{\partial \bar{\mathbf{H}}(\mathbf{r},\omega)}{\partial \omega} \cdot \mathbf{B}(\mathbf{r},\omega) \right]$$
$$-j\omega \dfrac{\partial \bar{\mathbf{H}}(\mathbf{r},\omega)}{\partial \omega} \cdot \mathbf{B}(\mathbf{r},\omega) - j\omega \dfrac{\partial \mathbf{E}(\mathbf{r},\omega)}{\partial \omega} \cdot \bar{\mathbf{D}}(\mathbf{r},\omega).$$

Making use of the first equation of (30) in the real frequency domain, we have

$$j\mathbf{E}(\mathbf{r},\omega) \cdot \bar{\mathbf{D}}(\mathbf{r},\omega) + j\bar{\mathbf{H}}(\mathbf{r},\omega) \cdot \mathbf{B}(\mathbf{r},\omega)$$
$$= \mathbf{E}(\mathbf{r},\omega) \cdot \dfrac{\partial \bar{\mathbf{J}}(\mathbf{r},\omega)}{\partial \omega} - \dfrac{\partial \mathbf{E}(\mathbf{r},\omega)}{\partial \omega} \cdot \bar{\mathbf{J}}(\mathbf{r},\omega)$$
$$- \left\{ \mathbf{E}(\mathbf{r},\omega) \cdot \nabla \times \dfrac{\partial \bar{\mathbf{H}}(\mathbf{r},\omega)}{\partial \omega} + \dfrac{\partial \bar{\mathbf{H}}(\mathbf{r},\omega)}{\partial \omega} \cdot \nabla \times \mathbf{E}(\mathbf{r},\omega) \right\}$$
$$+ \left\{ \dfrac{\partial \mathbf{E}(\mathbf{r},\omega)}{\partial \omega} \cdot \nabla \times \bar{\mathbf{H}}(\mathbf{r},\omega) - \bar{\mathbf{H}}(\mathbf{r},\omega) \cdot \nabla \times \dfrac{\partial \mathbf{E}(\mathbf{r},\omega)}{\partial \omega} \right\}$$
$$-j\omega \left[ \mathbf{E}(\mathbf{r},\omega) \cdot \dfrac{\partial \bar{\mathbf{D}}(\mathbf{r},\omega)}{\partial \omega} - \dfrac{\partial \mathbf{E}(\mathbf{r},\omega)}{\partial \omega} \cdot \bar{\mathbf{D}}(\mathbf{r},\omega) \right]$$
$$-j\omega \left[ \bar{\mathbf{H}}(\mathbf{r},\omega) \cdot \dfrac{\partial \mathbf{B}(\mathbf{r},\omega)}{\partial \omega} - \dfrac{\partial \bar{\mathbf{H}}(\mathbf{r},\omega)}{\partial \omega} \cdot \mathbf{B}(\mathbf{r},\omega) \right].$$

The terms inside the curved brackets can be expressed as a divergence of a vector with the aid of the vector identity $\nabla \cdot (\mathbf{a} \times \mathbf{b}) = \mathbf{b} \cdot \nabla \times \mathbf{a} - \mathbf{a} \cdot \nabla \times \mathbf{b}$. Thus

$$j\mathbf{E}(\mathbf{r},\omega) \cdot \bar{\mathbf{D}}(\mathbf{r},\omega) + j\bar{\mathbf{H}}(\mathbf{r},\omega) \cdot \mathbf{B}(\mathbf{r},\omega)$$
$$= \mathbf{E}(\mathbf{r},\omega) \cdot \dfrac{\partial \bar{\mathbf{J}}(\mathbf{r},\omega)}{\partial \omega} - \dfrac{\partial \mathbf{E}(\mathbf{r},\omega)}{\partial \omega} \cdot \bar{\mathbf{J}}(\mathbf{r},\omega)$$
$$+ \nabla \cdot \left( \mathbf{E}(\mathbf{r},\omega) \times \dfrac{\partial \bar{\mathbf{H}}(\mathbf{r},\omega)}{\partial \omega} - \dfrac{\partial \mathbf{E}(\mathbf{r},\omega)}{\partial \omega} \times \bar{\mathbf{H}}(\mathbf{r},\omega) \right)$$
$$+ j\omega \left[ \mathbf{B}(\mathbf{r},\omega) \dfrac{\partial \bar{\mathbf{H}}(\mathbf{r},\omega)}{\partial \omega} - \bar{\mathbf{H}}(\mathbf{r},\omega) \cdot \dfrac{\partial \mathbf{B}(\mathbf{r},\omega)}{\partial \omega} \right]$$
$$+ j\omega \left[ \bar{\mathbf{D}}(\mathbf{r},\omega) \cdot \dfrac{\partial \mathbf{E}(\mathbf{r},\omega)}{\partial \omega} - \mathbf{E}(\mathbf{r},\omega) \cdot \dfrac{\partial \bar{\mathbf{D}}(\mathbf{r},\omega)}{\partial \omega} \right].$$

Rearranging terms gives the new energy conservation law (36) in general materials.

## Acknowledgments

This work was supported in part by the Priority Academic Program Development of Jiangsu Higher Education Institutions and in part by the Jiangsu Innovation & Entrepreneurship Group Talents Plan.